\documentclass[12pt]{iopart}

\usepackage{iopams}
\usepackage{graphicx,amssymb}
\usepackage{color}
\newtheorem{lemma}{Lemma}
\newtheorem{proposition}{Proposition}
\newtheorem{theorem}{Theorem}

\newtheorem{definition}{Definition}

\newcommand{\be}{\begin{equation}}
\newcommand{\ee}{\end{equation}}
\def\dif{{\rm d}}

\begin{document}

\title[Kinematic approach to the classical ideal gas]
{Relativistic kinematic approach to the classical ideal gas}

\author{Joan Josep Ferrando$^{1,2}$ and Juan Antonio S\'aez$^3$}

\address{$^1$\ Departament d'Astronomia i Astrof\'{\i}sica, Universitat
de Val\`encia, E-46100 Burjassot, Val\`encia, Spain}

\address{$^2$\ Observatori Astron\`omic, Universitat
de Val\`encia, E-46980 Paterna, Val\`encia, Spain}

\address{$^3$\ Departament de Matem\`atiques per a l'Economia i l'Empresa,
Universitat de Val\`encia, E-46022 Val\`encia, Spain}

\ead{joan.ferrando@uv.es; juan.a.saez@uv.es}

\begin{abstract}
The necessary and sufficient conditions for a unit time-like vector field to be the unit velocity of a classical ideal gas are obtained. In a recent paper [Coll, Ferrando and S\'aez, Phys. Rev D {\bf 99} (2019)] we have offered a purely hydrodynamic description of a classical ideal gas. Here we take one more step in reducing the number of variables necessary to characterize these media by showing that a plainly kinematic description can be obtained. We apply the results to obtain test solutions to the hydrodynamic equation that model the evolution in local thermal equilibrium of a classical ideal gas.
\end{abstract}
%

\pacs{04.20.-q, 04.20.Jb}
%

%\maketitle

\section{Introduction}
\label{sec-intro}

In Relativity, a conservative energy tensor of the form $T = (\rho+ p) u \otimes u + p  g$ represents the energetic description of the evolution of a perfect fluid. If we want to describe the evolution of a perfect fluid in {\em local thermal equilibrium} we must add to the {\em hydrodynamic quantities}  ({\em unit velocity} $u$, {\em energy density} $\rho$, and {\em
pressure} $p$) a set of {\em thermodynamic quantities} ({\em matter density} $n$, {\em specific internal energy} $\epsilon$, {\em temperature} $\Theta$, and {\em specific entropy} $s$) constrained by the usual thermodynamic laws. This approach leads to a differential system:
\be  \label{H}
{\cal D}(u, \rho, p, n, \epsilon, s, \Theta) =0 \, ,
\ee
which is named the {\em fundamental system of the perfect fluid hydrodynamics}.

Elsewhere \cite{Coll-Ferrando-termo} \cite{CFS-LTE} we have shown that the system (\ref{H}) admits a {\em conditional system} for the hydrodynamic quantities $\{u, \rho, p\}$:
\be  \label{H-C}
{\tilde{\cal D}}(u, \rho, p) =0 \, .
\ee
This means that (\ref{H-C}) is a consequence of (\ref{H}), and conversely, for any solution $\{u, \rho, p\}$ of (\ref{H-C}), a solution $\{u, \rho, p, n, \epsilon, s, \Theta\}$ of (\ref{H}) exists. In other words, (\ref{H-C}) is the integrability condition for the system (\ref{H}) to admit a solution $\{n, \epsilon, s, \Theta\}$.

In \cite{CFS-LTE} we have analyzed in depth: (i) the {\em direct problem}, namely, the determination of the conditional system (\ref{H-C}) from the initial one (\ref{H}), and (ii) the {\em inverse problem}, namely, the obtainment of the solutions of (\ref{H}) associated with a given solution of (\ref{H-C}).

If we substitute in the fundamental system (\ref{H}) a generic equation of state for a particular one, corresponding to a specific perfect fluid, we can state the {\em restricted inverse and direct problems}. In \cite{CFS-LTE} we have solved these problems for the set of {\em generic ideal gases}, and this study has been applied to physically interpret some already known perfect fluid solutions of the Einstein equation \cite{C-F} \cite{CFS-CC} \cite{CFS-parabolic}.

Recently \cite{CFS-CIG}, we have performed a similar study for the {\em classical ideal gas} (CIG). We have solved the restricted direct problem by obtaining the conditional system in the hydrodynamic quantities (\ref{H-C}) associated with the {\em fundamental system of the classical ideal gas hydrodynamics} (\ref{H}). Thus, we have built a purely hydrodynamic description of the CIG.

Is it possible to build a purely kinematic approach to the CIG? More precisely, is it possible to express, solely in terms of the unit velocity $u$ and its derivatives, the necessary and sufficient conditions for $u$ to be the velocity of a classical ideal gas? The main goal of this paper is to show that the answer is affirmative: by starting from the hydrodynamic characterization (\ref{H-C}) we obtain a conditional system in the kinematic quantity $u$:
\be  \label{H-CC}
{\hat{\cal D}}(u) =0 \, ,
\ee
This result solves the restricted direct problem and offers a purely kinematic description of the CIG.

The search for the conditions in $u$ leads to a classification of the time-like unit vectors in eight classes. For each class, we obtain the necessary and sufficient conditions in $u$ to ensure that it is the velocity of some CIG. Furthermore, for each class we solve the inverse problem by obtaining the pairs $(\rho,p)$ that complete a solution to the system (\ref{H-C}).

It should be noted that a similar approach was carried out years ago for the {\em fundamental system of the barotropic hydrodynamic} \cite{CF-barotrop}. Furthermore, the study of conditional systems associated with a differential system and the analysis of the corresponding direct and inverse problems have shown their usefulness in other contexts. Thus, the Rainich \cite{rainich} theory for the non-null electromagnetic field precisely consists of
obtaining the conditional system for the Einstein-Maxwell energy tensor associated with Maxwell equations for the electromagnetic field. Also the Mariot-Robinson \cite{mariot} \cite{robinson} theorem on the null electromagnetic field imposes conditional constraints on its principal null direction.  Still in the electromagnetic framework we can quote the interpretation of the Teukolsky-Press relations \cite{teukolsky-press} \cite{CF-TP}, and in a more formal context the study of the Rainich approach to the Killing-Yano tensors \cite{fsKY} and to the Killing and conformal tensors \cite{cfs-KT}. The acquisition of IDEAL (intrinsic, deductive, explicit and algorithmic) characterizations of a metric or a family of metrics can also be formally identified as the answer to a direct problem (see %\cite{
%fsS, fsKerr,
\cite{fswarped} \cite{fs-SSST} \cite{fs-SS} \cite{Khavkine} \cite{Khavkine-b} and references therein).

This paper is organized as follows. In Section \ref{sec-CIG} we present the basic notation and concepts and we summarize the main results on the hydrodynamic approach to a classical ideal gas acquired in \cite{CFS-CIG}, which are the starting point for the present work.

Section \ref{sec-opening} is devoted to studying some generic constraints on the velocities of a classical ideal gas, and to analyzing the conditions that affect the richness of solutions of the inverse problem.

In section \ref{sec-classes} we characterize the velocities of a CIG. This study requires analyzing eight classes of unit vectors $u$. For every class, we offer the necessary and sufficient conditions for $u$ to be the velocity of a CIG, and we explain the full set of pairs $(\rho,p)$ which solve the inverse problem.

In section \ref{sec-quadres} we summarize the main results of the paper in an enlightening form. We present three tables. The first one offers the conditions that define the eight classes of unit vectors. The second one gives, for every class, the necessary and sufficient conditions charactering the CIG velocities. And the third one displays, for every class, the corresponding solution to the inverse problem. We also present a flow diagram with an algorithm enabling us to distinguish every class.

Section \ref{sec-CIG-examples} is devoted to presenting several examples of solutions to the fundamental system of the classical ideal gas hydrodynamics. In a first step we impose some significant constraints on the fluid flow (stationary, conformally stationary, geodesic) and we analyze the complementary conditions for this flow to be that of a classical ideal gas. And further, we solve the inverse problem to obtain the hydrodynamic quantities $(\rho,p)$ that complete the CIG solution of the hydrodynamic system.

%Finally, in section \ref{sec-remarks} we point out the interest of our results and comment about further extensions and applications in progress.

%\vspace{0.2cm}

In this paper we work on an oriented spacetime with a metric tensor
$g$ of signature $\{-,+,+,+\}$. For the metric product of two vectors, we write
$(x,y) = g(x,y)$, and we put $x^2 = g(x,x)$. The symbols $\nabla$, $\nabla \cdot$, $\dif$ and $*$ denote, respectively, the covariant derivative, the divergence operator, the exterior derivative and the Hodge dual operator, and $i(x)t$ denotes the interior product of a vector field $x$ and a p-form $t$.

%%%%%%%%%%%%%%%%%

\section{Classical ideal gas: hydrodynamic approach}
\label{sec-CIG}

The energetic description of the evolution of a perfect fluid is
given by its energy tensor:
\begin{equation}
\label{perfect-energy-tensor} T = (\rho+ p) u \otimes u + p \, g \, ,
\end{equation}
where $\rho$, $p$ and $u$ are, respectively,  the {\em energy
density}, {\em pressure} and {\em unit velocity} of the fluid.
A divergence-free $T$, $\nabla \cdot T = 0$, of this form is called
{\em perfect energy tensor}. These conservation equations take the
expression:
\begin{eqnarray}
\dif p  + \dot{p} u + (\rho + p) a = 0 \, ,  \label{con-eq1} \\[2mm]
\dot{\rho} + (\rho+ p) \theta = 0 \, ,   \label{con-eq2}
\end{eqnarray}
where $a=a[u]$ and $\theta=\theta[u]$ are, respectively, the acceleration and the
expansion of $u$, and where a dot denotes the directional
derivative, with respect to $u$, of a quantity $q$, $\dot{q} = u(q)
= u^{\alpha} \partial_{\alpha} q$. From now on, we write $h=h[u]$ to indicate that $h$ is a (tensorial) differential concomitant of the vector $u$.

A {\em barotropic evolution} is an evolution along which the {\em
barotropic relation} $\dif \rho \wedge \dif p = 0$ is fulfilled. A perfect
energy tensor describing energetically a barotropic evolution is
called a {\em barotropic perfect energy tensor}.

A perfect energy tensor $T$ represents the evolution in {\em local thermal equilibrium} (l.t.e.) of a perfect fluid if an associated {\em thermodynamic scheme} exists. This one can be obtained as the adiabatic
and Pascalian restriction of the Eckart's approach  \cite{Eckart},
and it implies introducing, besides the {\em hydrodynamic quantities} $\{u,\rho,p\}$, the {\em thermodynamic ones} $\{n, \epsilon, s, \Theta\}$. The {\em matter density} $n$, the {\em specific internal energy} $\epsilon$, the {\em
temperature} $\Theta$, and the {\em specific entropy} $s$, are submitted to: (i)  the decomposition
\begin{equation}
\rho= n(1+\epsilon) \, ,  \label{masa-energia}
\end{equation}
(ii) the conservation of matter:
\begin{equation}
\nabla \cdot (nu) = \dot{n} + n \theta = 0 \, .  \label{c-masa}
\end{equation}
and (iii) the {\em local thermal equilibrium
equation}:
\begin{equation}
\Theta \dif s = \dif \epsilon + p \dif(1/n) = (1/n) \dif \rho + (\rho+p) \dif (1/n)
\, .   \label{re-termo}
\end{equation}

We have already shown \cite{Coll-Ferrando-termo}  (see also the recent paper \cite{CFS-LTE})
that the notion of l.t.e admits a purely hydrodynamic formulation:
{\em a perfect energy tensor} $T$ {\em evolves in l.t.e if, and only
if, the hydrodynamic quantities $\{u,\rho,p\}$ fulfill the hydrodynamic sonic condition}
\begin{equation}
(\dot{\rho} \dif \dot{p} - \dot{p} \dif \dot{\rho}) \wedge \dif \rho \wedge \dif p
= 0     \, .       \label{h-lte}
\end{equation}
When the perfect energy tensor is non isoenergetic,
$\dot{\rho}\not=0$, condition (\ref{h-lte}) states that the
space-time function $\chi \equiv \dot{p}/\dot{\rho}$, called {\em
indicatrix of local thermal equilibrium}, depends only on the
quantities $p$ and $\rho$, $\chi = \chi (p,\rho )$. Then, this
function of state represents physically the square of the {\em speed
of sound} in the fluid, $\chi (\rho ,p) \equiv  c^2_{s}$.

The set of equations \{(\ref{con-eq1}),(\ref{con-eq2}),(\ref{masa-energia}),(\ref{c-masa}),(\ref{re-termo})\} constitutes the fundamental system of the perfect fluid hydrodynamics that has been expressed as (\ref{H}) in the introduction. And the set of equations \{(\ref{con-eq1}),(\ref{con-eq2}),(\ref{h-lte})\} is its associated conditional system expressed as (\ref{H-C}) in the introduction. Thus, the above quoted result solves the {\em generic
direct problem}, namely, the determination of the perfect energy tensors $T$ that model the l.t.e. evolution of any perfect fluid.

In practice, solving a {\em restricted direct problem} may be more
interesting than solving the generic one. In this way we have solved
in \cite{CFS-CIG} the direct problem for the family of classical ideal gases, which is defined by the equations of state:
\begin{equation}
p = k n \Theta > 0 \, , \quad     k \equiv {k_B \over m} \, , \quad \quad  \epsilon = c_v \Theta  \, ,
\label{CIG}
\end{equation}
$c_v > 0$ being the {\em heat capacity at constant volume}. Then, one obtains that a CIG has the
characteristic equation
\begin{equation}
\epsilon = \epsilon (n, s) =  \bar{\beta}(s) \, n^{\gamma -1}  ,
\quad \bar{\beta}(s) \equiv \exp{\frac{s-\bar{s}_0}{c_v}}  \, .
\label{epsilon-r-s}
\end{equation}
Moreover, any CIG satisfies the {\em classical $\gamma$-law}:
\begin{equation}
p=(\gamma-1)n \epsilon  , \qquad \gamma \equiv  1 + {k \over
c_v} > 1  \, ,   \label{g-law}
\end{equation}
$\gamma$ being the {\em adiabatic index}, and any
CIG fulfills a {\em Poisson law}:
\begin{equation}
p = \beta (s) n^{\gamma}   , \quad \beta(s) = (\gamma-1)
\bar{\beta}(s)    \, .               \label{poisson}
\end{equation}

We know \cite{CFS-LTE} that the only intrinsically barotropic ideal
gases are those satisfying $\epsilon(\Theta) = c_v \Theta -1$. Thus
CIG are, necessarily, non barotropic, $d \rho \wedge d p \not=0$,
and then we can take the hydrodynamic quantities $(\rho, p)$ as
coordinates in the thermodynamic plane, and we can obtain all the thermodynamic quantities in terms of them \cite{CFS-CIG}:
\begin{lemma}  \label{lemma-gic}
In terms of the hydrodynamic quantities $(\rho,p)$, the matter
density $n$, the specific internal energy $\epsilon$, the specific
entropy $s$ and the speed of the sound $c_s$ of a classical ideal
gas are given by
\begin{eqnarray}
n(\rho,p) = \rho - {p \over \gamma -1}   \, , \quad
\epsilon(\rho, p) = {p \over (\gamma -1) \rho - p}    \, , \qquad \label{r-e-cig} \\[2mm]
s(\rho,p) = s_0 + c_v \ln{p \over [\rho(\gamma -1) -p]^{\gamma}} \,
,   \qquad  \label{s-cig} \\[2mm]
c_s^2 = \chi(\rho,p) \equiv  {\gamma p \over \rho+p}  \, .  \qquad  \label{so-cig}
\end{eqnarray}
\end{lemma}

The solutions to the direct problem for the classical ideal gas and the specific inverse problem obtained in \cite{CFS-CIG} can be summarized in the following two statements:
\begin{proposition} \label{propo-iso}
The necessary and sufficient condition for a non barotropic and
isoenergetic ($\dot{\rho} =0$) perfect energy tensor $T=(u,\rho,p)$
to represent the l.t.e. evolution of a CIG is to be
isobaroenergetic: $\dot{\rho}=0$, $\dot{p}=0$. Then $T$ represents
the evolution in l.t.e. of any CIG, and the specific internal energy
$\epsilon$, the matter density $n$, the specific entropy $s$ and the
speed of sound $c_s$ are given by {\em (\ref{r-e-cig}),
(\ref{s-cig})} and {\em (\ref{so-cig})}.
\end{proposition}

\begin{proposition} \label{propo-CIG}
The necessary and sufficient condition for a non barotropic and non
isoenergetic perfect energy tensor $T=(u,\rho,p)$ to represent the
l.t.e. evolution of a classical ideal gas with adiabatic index
$\gamma$ is that the indicatrix function $\chi$ be of the form:
\begin{equation}
\chi \equiv \frac{\dot{p}}{\dot{\rho}} = \frac{\gamma p}{\rho + p}  \, .      \label{chi-CIG}
\end{equation}
Then, the matter density $n$ and the specific entropy $s$ are given
by {\em (\ref{r-e-cig})} and {\em (\ref{s-cig})}, and the constants
$k$ and $c_v$ are related by {\em (\ref{g-law})}.
\end{proposition}

Note that the set of equations \{(\ref{con-eq1}),(\ref{con-eq2}),(\ref{masa-energia}),(\ref{c-masa}),(\ref{CIG})\} constitutes the fundamental system of the CIG hydrodynamics that has been stated as (\ref{H}) in the introduction. And the set of equations \{(\ref{con-eq1}),(\ref{con-eq2}),(\ref{chi-CIG})\} is its associated conditional system that has been stated as (\ref{H-C}) in the introduction. Thus, the above propositions \ref{propo-iso} and \ref{propo-CIG} solve the  restricted direct and inverse problems, namely, the determination of the perfect energy tensors $T$ that model the l.t.e. evolution of a CIG, and the acquisition of the full set of thermodynamic quantities associated with them.

Although a classical ideal gas can not be intrinsically barotropic, it may have a barotropic evolution when a determined function of state remains constant in this evolution. When this function of state is not the specific entropy $s$, this evolution is, necessarily, isobaroenergetic, $\dot{\rho}= \dot{p} = 0$ \cite{CFS-LTE} \cite{CFS-CIG}. And when the evolution is isentropic, one has a specific barotropic relation $p= \phi(\rho)$ \cite{CFS-CIG}. More precisely we have:
\begin{proposition}  \label{propo-bar-CIG}
A barotropic perfect energy tensor $T=(u,\rho,p)$ represents the evolution of a CIG if one of the two following condition holds:
\begin{itemize}
\item[(i)]
It is isobaroenergetic: $\dot{\rho}= \dot{p} = 0$. Then, $T$ represents the evolution of any CIG.
\item[(ii)]
It is an isentropic evolution of a CIG with adiabatic index $\gamma$, and the barotropic relation is
\be \label{rop-isentropic}
(\gamma-1) \rho = p + B p^{\frac{1}{\gamma}} \, , \quad B= {\rm constant} \not=0 \, .
\ee
\end{itemize}
\end{proposition}

In \cite{CFS-CIG} we have shown that  both, the {\em $\gamma$-gases} defined by the classical $\gamma$-law (\ref{g-law}) and the {\em Poisson gases} defined by the Poisson law (\ref{poisson}), have the expression (\ref{so-cig}) for the square of the speed of sound. This means that these media admit the same perfect energy tensors $T$ as solution to the direct problem. Consequently, our study of the CIG velocities undertaken in this paper is also valid for these media. In fact, in this paper we characterize the velocities of the Poisson gases, which include the $\gamma$-gases and the classical ideal gases. Nevertheless, it is worth remarking that the solution of the inverse problem is wider for these media \cite{CFS-CIG}: the specific entropy is an arbitrary function of (\ref{s-cig}) and, for the Poisson gases, the expressions of the matter density and the internal energy involve another arbitrary function of (\ref{s-cig}). Thus, the expressions (\ref{r-e-cig}) and (\ref{s-cig}) for the thermodynamic quantities are only valid for the CIG.

%%%%%%%%%%%%%%%%%%%%%%%%%%

\section{Velocities of a classical ideal gas. Opening results}
\label{sec-opening}

For a CIG, an isoenergetic evolution is equivalent to $\theta=0$ as a consequence of (\ref{con-eq2}) ($\rho + p>0$). Thus, proposition \ref{propo-CIG} offers the hydrodynamic characterization of a CIG when $\theta \not=0$, that is, the hydrodynamic quantities $\{u, \rho, p\}$ are submitted to system (\ref{con-eq1}) (\ref{con-eq2}) and (\ref{chi-CIG}). The last equation can be replaced by:
\be \label{dotp}
\dot{p} + \gamma \theta p = 0 \, .
\ee
Then, (\ref{con-eq1}) becomes:
\be \label{dlnp}
\dif \ln p^{1/\gamma} = \theta u - f a \, , \qquad f \equiv \frac{1}{\gamma}\left[\frac{\rho}{p} + 1\right] \not = 0  \, .
\ee
Moreover, from this definition of $f$ and from (\ref{con-eq2}) and (\ref{dotp}) we obtain:
\be \label{fdot}
\dot{f} = \theta [(\gamma-1) f - 1] \, .
\ee
And, from the first expression in (\ref{dlnp}),
\be \label{dfa}
\dif (\theta u - f a) = 0 \, .
\ee
Conversely, if a function $f \not=0$ that fulfills conditions (\ref{fdot}) and (\ref{dfa}) exists, then we can find a function $p$ submitted to the first expression in (\ref{dlnp}), and consequently to (\ref{dotp}). Moreover, if we define $\rho = p(\gamma f - 1)$, then (\ref{dlnp}) and (\ref{fdot}) imply (\ref{con-eq1}) and (\ref{con-eq2}). Note that the solutions with a constant $f$ are forbidden when $\theta\not=0$ since they lead to a barotropic evolution of the form $p=(\gamma-1) \rho$, which is only compatible with an isobaroenergetic energy tensor, that is, $\theta=0$ (see proposition \ref{propo-bar-CIG}). Thus, we have proved:
\begin{proposition}  \label{propo-uf-gamma}
The necessary and sufficient condition for a time-like unit vector field $u$ with $\theta \not=0$ to be the velocity of a CIG with adiabatic index $\gamma > 1$ is that a non constant function $f$ exists such that the pair $(u,f)$ fulfills equations {\em (\ref{fdot})} and {\em (\ref{dfa})}.
\end{proposition}

Condition (\ref{fdot}) in the above proposition involves the adiabatic index $\gamma$. We can also obtain a characterization which is valid for any CIG, that is, not involving a previously fixed $\gamma$:
\begin{proposition} \label{propo-uf}
The necessary and sufficient condition for a time-like unit vector field $u$ with $\theta \not=0$ to be the velocity of a CIG  is that a non constant function $f$ exists such that the pair $(u,f)$ fulfills equation {\em (\ref{dfa})} and
\be  \label{dg}
\tilde{\gamma} > 0 \, , \qquad  d \tilde{\gamma} = 0 \, , \qquad \tilde{\gamma} \equiv \frac{1}{f} \left(1+\frac{\dot{f}}{\theta}\right) \, .
\ee
\end{proposition}
The determination of the hydrodynamic quantities $(\rho,p)$ from the pair $(u,f)$ is as follows:
\begin{proposition} \label{propo-uf-rop}
If a pair $(u, f)$ fulfills conditions {\em (\ref{dfa})} and {\em (\ref{dg})} in proposition {\em \ref{propo-uf}}, then a function $\psi$ and a constant $\gamma$ exist such that
\be
 \dif \psi = \theta u - f a \, , \qquad \gamma = \tilde{\gamma} +1  \, .
\ee
Then, $u$ is the velocity of a classical ideal gas with adiabatic index $\gamma$, and the pressure and the energy density are given, respectively, by
\be \label{pro}
 p = C e^{\gamma \psi}  \, , \qquad \rho = p(\gamma f -1)  \, ,
\ee
where $C$ is a constant.
\end{proposition}

Note that the pair $(u,f)$ determines the pair $(\rho,p)$ up to a constant factor $C$. This invariance $(\rho, p) \rightarrow (C \rho, C p)$ can be inferred from the initial equations (\ref{con-eq1}), (\ref{con-eq2}) and (\ref{chi-CIG}) for a given velocity $u$.

In order to characterize the velocities of a CIG we must find the conditional system in $u$ for the differential system \{(\ref{fdot}),(\ref{dfa})\} in $(u,f)$. Before dealing with this problem in the next section it is worth analyzing two questions that naturally arise from the statements in propositions \ref{propo-uf-gamma}, \ref{propo-uf} and \ref{propo-uf-rop}.

The first one states: If $(u,f_0)$ is a solution of the system \{(\ref{dfa}),(\ref{dg})\}, is there another $f \not= f_0$ such that $(u, f)$ is also a solution? Note that if both $f_0$ and $f$ fulfill equation (\ref{dfa}), we obtain
\be \label{dffa}
\dif  [(f-f_0)a]=0 \, ,
\ee
and, consequently, when acceleration $a$ does not vanish, it defines an integrable one-form, $a \wedge \dif a=0$. The case $a=0$ will be considered in next section. If $a\not=0$, we have $f=f_0+ \beta \varphi(\alpha)$, where $\alpha$ and $\beta$ are, respectively, an integrant factor and a potential of $a$, $\beta a = \dif \alpha$. Thus, we have proved:

\begin{lemma} \label{lemma-dues-f}
Let $u$ be a non-geodesic ($a \not=0$) and expanding ($\theta\not=0$) unit vector. A necessary condition for the differential system {\em \{(\ref{fdot}),(\ref{dfa})\}} to admit two solutions $(u,f_0)$ and $(u,f)$, $f \not= f_0$, is that $a \wedge \dif a = 0$. Moreover, if so, $f=f_0+ \beta \varphi(\alpha)$, where $\alpha$ and $\beta$ are, respectively, an integrant factor and a potential of $a$, $\beta a = \dif \alpha$.
\end{lemma}

Note that the vector $v= *(a \wedge \dif a)$ measures the 'vorticity' of the acceleration vector $a$. Thus, the necessary condition in lemma \ref{lemma-dues-f} states that acceleration $a$ is a hypersurfice-orthogonal vector, $v=0$. In next section we obtain a necessary and sufficient condition for the existence of more than one solution $f$ (see proposition \ref{propo-dues-f}).

The second question states: if $(u,f)$ is a solution of the system \{(\ref{fdot}),(\ref{dfa})\}, what additional conditions guarantee a barotropic evolution? Evidently, (\ref{pro}) implies that $\dif \rho \wedge \dif p =0$ is equivalent to $\dif f \wedge  \dif p = 0$, and (\ref{dlnp}) allows us to write this condition in terms of the variables $(u,f)$. Moreover, the evolution is, necessarily, isentropic as a consequence of proposition \ref{propo-bar-CIG}. Consequently, we obtain:
\begin{proposition} \label{propo-isentropic}
A pair $(u, f)$ that fulfills the conditions in proposition {\em \ref{propo-uf}} (or {\em \ref{propo-uf-gamma})} defines a barotropic (and then isentropic) evolution of a CIG if, and only if,
\be \label{isentropic}
\dif f \wedge (\theta u - f a) = 0 \, .
\ee
\end{proposition}

It is worth remarking that under the barotropic constraint (\ref{isentropic}), from (\ref{fdot}) and (\ref{pro}) we obtain $\rho'(p) = f = \frac{1}{\gamma}(\frac{\rho}{p} +1)$. This equation can be integrated and leads to a solution of the form (\ref{rop-isentropic}), accordingly with the statement of proposition above.
%
%%%%%%%%%%%%%%%%%%%%%%%%%%
%
\section{Velocities of a classical ideal gas. Classes and characterization}
\label{sec-classes}

The study of the conditional system in $u$ associated with the differential system \{(\ref{con-eq1}),(\ref{con-eq2}),(\ref{chi-CIG})\} leads to a classification of the time-like unit vector fields. For each class, we must obtain the necessary and sufficient conditions on $u$ and its differential concomitants to ensure that $u$ is the velocity of a CIG, and we must give the richness of pairs $(\rho, p)$ that complete the solution.

\subsection{Case $\theta =0$}
\label{subsec-theta=0}

When $\theta=0$ the evolution is isobaroenergetic, $\dot{\rho}=\dot{p}=0$, and $d p = - (\rho+p) a$. These conditions imply:
\be  \label{theta=0}
a \wedge \dif a = 0 \, , \qquad i(u)\dif a = 0 \, .
\ee
Conversely, if (\ref{theta=0}) holds, we can find pairs $(\rho, p)$ that fulfill the conservation equations. More precisely, if we distinguish the case of geodesic motion, we easily obtain:

\begin{proposition} \label{propo-theta=0}
An expansion-free ($\theta =0$) time-like unit vector $u$ is the velocity of a CIG if, and only if, it fulfills equations {\em (\ref{theta=0})}.\\
If $u$ is geodesic ($a=0$), the pressure is an arbitrary constant, $p=p_0$, and the energy density is an arbitrary $u$-invariant function, $\rho= \rho(\varphi_i)$, $\dot{\varphi_i}=0$, $i=1,2,3$.\\
If $u$ is not geodesic ($a \not=0$), the pressure in an arbitrary potential of $a$, $p=p(\alpha)$, $\dif \alpha = \beta a$, and the energy density is given by $\rho = - p - \frac{1}{a^2}(d p, a)= - p(\alpha) - \beta p'(\alpha)$.
\end{proposition}

Note that the barotropic case occurs when $\dif a=0$. Moreover, when $a=0$ the energy density $\rho = \rho(\varphi_i)$ is an arbitrary $u$-invariant function, that is, a function of three independent $u$-invariant functions $\varphi_i$. Thus, the dimension of the pairs $(\rho,p)$ associated with a given $u$ is controlled by an arbitrary constant and an arbitrary function of three variables when $a=0$, and by an arbitrary real function when $a \not=0$. We have $\dot{\rho}= \dot{p} =0$ in this case and, accordingly with proposition \ref{propo-iso}, CIG  with any adiabatic index $\gamma$ can be associated with $u$.
%
%%%%%%%%%%%%%%%%%%%%%%
%
\subsection{Case $\theta \not=0$, $a=0$}
\label{subsec-a=0}

If $a=0$, condition (\ref{dlnp}) implies $d \ln p^{1/\gamma} = \theta u$, and thus $\dif (\theta u) = 0$. Then, we have necessarily
\begin{equation}  \label{a=0}
\dif u = 0 \, , \qquad \dif \theta \wedge u = 0 \, .
\ee
Conversely, if conditions (\ref{a=0}) hold, a function $t$ exists such that
\be \label{a=0-t}
u = - dt \, , \qquad \theta = \theta(t) \, .
\ee
Then, we can obtain the pressure as
\be \label{a=0-p}
\displaystyle p = C e^{- \gamma \! \int \! \theta(t) dt} \, ,
\ee
and, from (\ref{chi-CIG}), the energy density $\rho$ is any solution to linear equation $\dot{\rho} = \frac{\dot{p}}{\gamma p}(\rho + p)$. For $\gamma \not=1$, $\rho$ is then of the form
\be  \label{a=0-rho}
(\gamma-1) \rho = p + B(\varphi_i) p^{\frac{1}{\gamma}} \, , \quad \dot{\varphi_i}=0, \ \   i=1,2,3 \, .
\ee
Note that $B(\varphi_i)$ is an arbitrary $u$-invariant function. Thus, we obtain:
\begin{proposition} \label{propo-a=0}
An expanding ($\theta \not=0$) and geodesic ($a=0$) time-like unit vector $u$ is the velocity of a CIG if, and only if, it fulfills equations {\em (\ref{a=0})}.\\
Then, a function $t$ exists such that relations {\em (\ref{a=0-t})} hold and, for every adiabatic index $\gamma$, the pressure and the energy density are given, respectively, by {\em (\ref{a=0-p})} and {\em (\ref{a=0-rho})}.
\end{proposition}

Note that the barotropic case occurs when $B(\varphi_i)=constant$, and then we have an isentropic evolution. Moreover, the dimension of the pairs $(\rho,p)$ associated with a given $u$ is controlled by an arbitrary constant $C$ and an arbitrary function of three variables $B(\varphi_i)$. Constraints (\ref{a=0}) do not restrict the adiabatic index $\gamma$. Consequently, any CIG can be associated with a given $u$.

%%%%%%%%%%%%%%%%%%%%%%

\subsection{Case $\theta a \not=0$, $(v, u)\not=0$}
\label{subsec-vunot=0}

When $\theta a \not=0$, proposition \ref{propo-uf-gamma} applies and we must look for a function $f$ such that the pair $(u,f)$ fulfill equations (\ref{dfa})  and (\ref{dg}). The first one is equivalent to:
\be \label{dfa-2}
\dif \theta \wedge u  + \theta \dif u - \dif f \wedge a - f \dif a = 0 \, .
\ee
The exterior product of this equation by $a$ leads to:
\be \label{du-a}
\dif \theta \wedge u \wedge a + \theta \dif u \wedge a = f \dif a \wedge a \, .
\ee
And the exterior product of this last equation by $u$ implies:
\be \label{du-a-u}
\theta \dif u \wedge a \wedge u = f \dif a \wedge a \wedge u \, .
\ee
Let us consider the vorticity of the fluid, $\omega = *(u \wedge \dif u)$, and the vorticity of the acceleration vector, $v = *(a \wedge \dif a)$. Then, if $(v,u) \not=0$, (\ref{du-a-u}) allows us to obtain $f$ as:
\be \label{f-vunot=0}
f = f_1[u] \equiv -\theta \frac{(\omega, a)}{(v, u)} \, .
\ee
Thus, we have:
\begin{proposition} \label{propo-vunot=0}
A time-like unit vector field $u$ with $\theta (v,u) \not=0$ is the velocity of a CIG if, and only if, the pair $(u,f)$, where $f$ is given in {\em (\ref{f-vunot=0})}, fulfills conditions in proposition {\em \ref{propo-uf}}.\\
Then, the adiabatic index $\gamma$, the pressure $p$ and the energy density $\rho$ are determined as stated in proposition {\em \ref{propo-uf-rop}}.
\end{proposition}

It is worth remarking that proposition \ref{propo-uf} bans a constant value for $f$ and, in particular, $f=0$ is forbidden. Thus $(\omega,a)\not=0$, and this class is not compatible with an irrotational motion.

Note that, as a consequence of proposition \ref{propo-isentropic}, the barotropic case occurs when (\ref{isentropic}) is also fulfilled. Moreover, the dimension of the pairs $(\rho,p)$ associated with a given $u$ is controlled by an arbitrary constant $C$. In this case the adiabatic index of the CIG is fixed by $u$.

%%%%%%%%%%%%%%%%%%%%%%

\subsection{Case $\theta a \not=0$, $(v, u)=0$, $v\not=0$}
\label{subsec-vu=0}

As $\theta \not=0$ proposition \ref{propo-uf-gamma} applies. And $(v,u)=0$ implies that $v$ is a space-like vector ($v^2 >0$). Moreover $\dif u \wedge a \wedge u = 0$ (($\omega,a)=0$) as a consequence of (\ref{du-a-u}). This last condition is equivalent to $\dif u \wedge a = 0$. Then, equation (\ref{du-a}) becomes:
\be
x = f v \, , \qquad  x = x[u] \equiv *(\dif \theta \wedge u \wedge a) \, ,
\ee
and, consequently, $f$ can be obtained as:
\be \label{f-vu=0}
f = f_2[u] \equiv \frac{(x, v)}{v^2} \, .
\ee
Thus, we have:
\begin{proposition} \label{propo-vu=0}
A time-like unit vector $u$ with $\theta v\not=0$ and $(v,u)=0$ is the velocity of a CIG if, and only if, the pair $(u,f)$, where $f$ is given in {\em (\ref{f-vu=0})}, fulfills conditions in proposition {\em \ref{propo-uf}}.\\
Then, the adiabatic index $\gamma$, the pressure $p$ and the energy density $\rho$ are determined as stated in proposition {\em \ref{propo-uf-rop}}.
\end{proposition}

It is worth remarking that proposition \ref{propo-uf} bans a constant value for $f$ and, in particular, $f=0$ is forbidden. Thus $x\not=0$, and in this class the expansion gradient $\dif \theta$ can not lie on the plane $\{u,a\}$.

Note that, as a consequence of proposition \ref{propo-isentropic}, the barotropic case occurs when (\ref{isentropic}) is also fulfilled. Moreover, the dimension of the pairs $(\rho,p)$ associated with a given $u$ is controlled by an arbitrary constant $C$. In this case the adiabatic index of the CIG is fixed by $u$.

%%%%%%%%%%%%%%%%%%%%%%

\subsection{Classes with $\theta a \not=0$, $v=0$}
\label{subsec-v=0}

Now, proposition \ref{propo-uf-gamma} applies again. Moreover, $v=0$ and (\ref{du-a}) imply $x=0$ and $\dif u \wedge a =0$. The first condition states that the gradient of the expansion lies on the plane $\{u,a\}$:
\be \label{dtheta}
\dif \theta = - \dot{\theta} u + \theta^{*} a \, , \qquad \theta^* = \frac{1}{a^2} i(a)\dif \theta \, .
\ee
And the second one implies that two vectors $b$ and $c$ exist such that
\begin{eqnarray} \label{b}
\dif u =a \wedge u + a \wedge b \, , \qquad \,  (b,u)=(b,a)=0 \, , \\[1mm]
\label{daca}
\dif a = a \wedge c \,  ,\quad \ \  (c,a)=0 \, , \quad  \ c = \frac{1}{a^2} i(a) \dif a \, .
\end{eqnarray}
Note that $\omega = *(u \wedge a \wedge b)$ and, consequently, the motion is vorticity-free ($\omega=0$) if, and only if, $b=0$. Thus, when $\omega \not=0$, $\{u,a,b,\omega\}$ is an orthogonal tetrad.

From relations (\ref{daca}), (\ref{dtheta}) and (\ref{b}), equation (\ref{dfa-2}) (equivalent to (\ref{dfa})) becomes
\be \label{dfa-3}
a \wedge [\dif f + (\theta + \theta^*) u + \theta b - fc] = 0  \, .
\ee
And the contraction of this equation with $u$ and $a$ leads to:
\be \label{dfa-4}
\dot{f} - f (c,u) =  \theta +  \theta^*  \, .
\ee
We can eliminate $\dot{f}$ by using this equation and (\ref{fdot}), and we obtain:
\be \label{XY}
f (\tilde{\gamma} - \mu) = \nu  \, ,
\ee
where
\be \label{XY-2}
\mu  = \mu[u] \equiv \frac{(c,u)}{\theta} \, , \qquad \nu = \nu[u] \equiv 2 + \frac{\theta^*}{\theta}  \, .
\ee
If we take the $u$-derivative of (\ref{XY}) and make use of (\ref{fdot}) again, we can eliminate $\dot{f}$ and obtain:
\be \label{XY-3}
\theta(f \tilde{\gamma}- 1)(\tilde{\gamma} - \mu) - f \dot{\mu}  = \dot{\nu}   \, .
\ee
Finally, we can eliminate $\gamma$ by using equations (\ref{XY}) and (\ref{XY-3}), and we obtain the following equation for $f$:
\be \label{XY-4}
f^2 \dot{\mu} + f( \dot{\nu} - \theta \mu \nu) + \theta \nu (1- \nu)  = 0  \, .
\ee
We can summarize the results in this subsection in the following.
\begin{lemma} \label{lemma}
If a time-like unit vector $u$ with $\theta a \not=0$ and $v=0$ is the velocity of a CIG, then function $f$ in proposition {\em \ref{propo-uf}} is submitted to equation {\em (\ref{XY-4})}. Moreover, relation {\em (\ref{XY})} is also fulfilled.
\end{lemma}
%

%%%%%%%%%%%%%%%%%%%%%%

\subsection{Case $\theta a \not=0$, $v=0$, $\nu \not=0$}
\label{subsec-nunot=0}

Under these constraints, equation (\ref{XY-4}) is a second degree equation for $f$ when $\dot{\mu} \not=0$. Then, we obtain real solutions if
\be \label{Delta}
\Delta = \Delta[u] \equiv  (\dot{\nu} - \theta \mu \nu)^2 - 4 \dot{\mu} \theta \nu (1- \nu) \geq 0 \, ,
\ee
and they are given by:
\be \label{f+-}
f = f_{\pm}[u] \equiv \frac{1}{2 \dot{\mu}}[\theta \mu \nu - \dot{\nu} \pm \sqrt{\Delta}] \, .
\ee
Otherwise, if $\dot{\mu} =0$, equation (\ref{XY-4}) admits a solution if $\dot{\nu} - \theta \mu \nu \not=0$:
\be \label{Xdot=0}
f = f_3[u] \equiv \frac{\theta \nu (1- \nu)}{\dot{\nu} - \theta \mu \nu} \,  .
\ee
Note that condition $\dot{\nu} - \theta \mu \nu =0$ implies $\nu=1$, and then $\mu=0$. Thus, from (\ref{XY}) we obtain $f = 1/\tilde{\gamma} = constant$, and $(u,f)$ does not satisfies conditions in proposition \ref{propo-uf}. Thus, we have:
\begin{proposition} \label{propo-v=0-Ynot0}
A time-like unit vector field $u$ with $\theta \nu a\not=0$, $v=0$ and $\dot{\mu}\not=0$ is the velocity of a CIG if, and only if, {\em (\ref{Delta})} holds and the pair $(u,f)$, where $f$ is one of the functions given in {\em (\ref{f+-})}, fulfills conditions in proposition {\em \ref{propo-uf}}.\\
A time-like unit vector $u$ with $\theta \nu a\not=0$, $v=0$ and $\dot{\mu}=0$ is the velocity of a CIG if, and only if, $\dot{\nu} - \theta \mu \nu  \not= 0$ and the pair $(u,f)$, where $f$ is given in {\em (\ref{Xdot=0})}, fulfills conditions in proposition {\em \ref{propo-uf}}.\\
In both cases, the adiabatic index $\gamma$, the pressure $p$ and the energy density $\rho$ are determined as stated in proposition {\em \ref{propo-uf-rop}}.
\end{proposition}

It is worth remarking that proposition \ref{propo-uf} bans a constant value for $f$ and, in particular, $f=0$ is forbidden. Thus, when $\dot{\mu}=0$ we have, necessarily, $\nu\not=1$.

Note that, as a consequence of proposition \ref{propo-isentropic}, the barotropic case occurs when (\ref{isentropic}) is also fulfilled. Moreover, the dimension of the pairs $(\rho,p)$ associated with a given $u$ is controlled by an arbitrary constant $C$.  In this case the adiabatic index of the CIG is fixed by $u$.

%%%%%%%%%%%%%%%%%%%%%%

\subsection{Case $\theta a \not=0$, $v=0$, $\nu =0$}
\label{subsec-nu=0}

In the previous cases with $\nu \not= 0$ the scalar $f$ is univocally determined by $u$. Thus, if $\theta a \not=0$, then $v=0$ and $\nu =0$ are necessary conditions for the existence of two different functions $f$ and $f_0$ which are solutions to the differential system (\ref{dfa}-\ref{dg}). But this conditions are also sufficient. Indeed, under these constraints, equation (\ref{XY}) implies $\mu = \tilde{\gamma}$ and the adiabatic index $\gamma = \tilde{\gamma} +1$ is the the same for both $f$ and $f_0$. Then, if $f=f_0+ \beta \varphi(\alpha)$, the pair $(u,f)$ fulfills conditions of proposition \ref{propo-uf} provided that $(u,f_0)$ fulfills. Consequently, we obtain:
\begin{proposition} \label{propo-dues-f}
Let $u$ be a non-geodesic ($a \not=0$) and expanding ($\theta\not=0$) unit vector. If the differential system {\em \{(\ref{fdot}),(\ref{dfa})\}} admits a solution $(u,f_0)$, then it admits another solution $(u,f)$ if, and only if, $u$ fulfills $v=0$ and $\nu=0$. Moreover,
\be  \label{f-f0}
f=f_0+ \beta \varphi(\alpha) \, ,
\ee
where $\alpha$ and $\beta$ are, respectively, an integrant factor and a potential of $a$, $\beta a = \dif \alpha$, and $\varphi(\alpha)$ an arbitrary real function.
\end{proposition}

We assume that a solution $f_0$ exists. Then, equation (\ref{XY-4}) is an identity, and from the definitions of $\mu$ and $\nu$ given in (\ref{XY-2}) we have:
\be \label{cu-theta}
(c,u) = \tilde{\gamma}\theta \, , \qquad  \theta^* + 2 \theta = 0 \, .
\ee
Then, (\ref{dtheta}) becomes:
\be \label{dtheta-2}
\dif \theta = - \dot{\theta} u -2 \theta a \, .
\ee
From (\ref{daca}), (\ref{b}), (\ref{cu-theta}) and the integrability condition of (\ref{dtheta-2}) we obtain:
\be \label{c-ub}
c= - \tilde{\gamma} \theta u - \frac{\dot{\theta}}{2 \theta} b \, , \qquad d \dot{\theta} \wedge u \wedge a = 0 \, .
\ee

First we study the case $\omega \not=0$. Then $b \not=0$, and from the expression (\ref{c-ub}) and the integrability conditions of (\ref{daca}) and (\ref{b}), we obtain:
\be \label{thetadots-thetadot}
\ddot{\theta} = \dot{\theta} \left[\frac{3\dot{\theta}}{2 \theta} -  \tilde{\gamma} \theta \right]  \,  .
\ee
Taking into account (\ref{dtheta-2}) and (\ref{c-ub}), $\dot{\theta}=0$ implies $\dif a=0$, that is $c=0$, and then $\tilde{\gamma}=0$. Thus, the case $\dot{\theta}=0$ is not compatible. If $\dot{\theta} \not=0$ we can define
\be \label{f0}
f_0 = f_4[u] \equiv - \frac{2 \theta^2}{\dot{\theta}} \, .
\ee
Then, we can easily prove that $f_0$ fulfills conditions in proposition \ref{propo-uf}.

When $\omega=0$, that is, $b=0$, (\ref{daca}) and (\ref{b}) become:
\be
\dif u = a \wedge u \, , \qquad \dif a =  - \theta \tilde{\gamma} a \wedge u \, .
\ee
Note that (\ref{cu-theta}) forbids the case $(c,u)=0$. Otherwise, we can define:
\be \label{f0-2}
f_0 = f_5[u] \equiv - \frac{\theta}{(c,u)} \, ,
\ee
that is, $f_0 = 1/\tilde{\gamma}$ fulfills conditions (\ref{dfa}) and (\ref{dg}) but it does not fulfill all the conditions in proposition \ref{propo-uf} because it is a constant. But the family of $f$ defined by (\ref{f-f0}) should be considered.

The two cases, $\omega=0$ and $\omega \not=0$ can be summarized in the following.
\begin{proposition} \label{propo-v=0-Y=0}
A time-like unit vector $u$ with $\theta  a \not=0$, $v=0$, $\nu=0$ and $\omega\not=0$ is the velocity of a CIG if, and only if, $\dot{\theta}\not=0$ and the pair $(u,f_0)$, where $f_0$ is given in {\em (\ref{f0})}, fulfills conditions {\em (\ref{dfa})} and {\em (\ref{dg})}.\\
A time-like unit vector $u$ with $\theta  a \not=0$, $v=0$, $\nu=0$ and $\omega=0$ is the velocity of a CIG if, and only if, $(c,u) \not=0$ and the pair $(u,f_0)$, where $f_0$ is given in {\em (\ref{f0-2})}, fulfills conditions {\em (\ref{dfa})} and {\em (\ref{dg})}.\\
In both cases, the non-constant functions of the form $f = f_0 +\beta \varphi(\alpha)$, $\beta a = \dif \alpha$, satisfy the conditions in proposition {\em \ref{propo-uf}} and, for every $f$, the adiabatic index $\gamma$, the pressure $p$ and the energy density $\rho$ are determined as stated in proposition {\em \ref{propo-uf-rop}}.
\end{proposition}

Note that, as a consequence of proposition \ref{propo-isentropic}, the barotropic case occurs when (\ref{isentropic}) is also fulfilled.  Moreover, the dimension of the pairs $(\rho,p)$ associated with a given $u$ is controlled by an arbitrary constant $C$ and an arbitrary real function $\varphi(\alpha)$. The adiabatic index of the CIG is fixed by $u$ and then it is the same for every the pair $(\rho,p)$.
%
%%%%%%%%%%%%%%%%%%%%%%

%%%%%%%%%%%%%%%%%%%%%%
%
\section{Velocities of a classical ideal gas: summary theorems}
\label{sec-quadres}

In the above section we have obtained conditional systems in $u$ for the  fundamental system of the CIG hydrodynamics. These systems are constraints imposed on some differential quantities associated with $u$ and they do not admit a unique simple form valid for any unit vector. On account of the above results, we are led to introduce the following classification of the unit vector fields.
%
%\vspace{-2mm}
\begin{table}[b]
\noindent
%\vspace{-2mm}
%\begin{ruledtabular}
\normalsize{
\begin{tabular}{cl}
\ \ \ Classes \ \ \  &\phantom{\Large $\frac{A}{B}$} Definition relations  \\
\hline \\[-5mm] \hline

C$_1$ & \phantom{\Large $\frac{A}{B}$}  $\theta =0$   \\
\hline

C$_2$ & \phantom{\Large $\frac{A}{B}$} $\theta \neq 0, \qquad \qquad \  a=0 $ \\
\hline

C$_3$ & \phantom{\Large $\frac{A}{B}$}  $\theta \,  a \neq 0  $, \qquad \quad \ $(v,u) \neq 0  $ \\
\hline

C$_4$ & \phantom{\Large $\frac{A}{B}$}  $\theta \, a \neq 0 $, \qquad \quad \  
$(v,u) = 0 $,\quad \qquad   $v \neq  0 $ \\ \hline

C$_5$ & \phantom{\Large $\frac{A}{B}$} $\theta \, a \neq 0$, \qquad \quad  \ $v=0 $, \qquad \qquad \ $\nu \, \dot{\mu} \neq 0 $ \\
\hline

C$_6$ & \phantom{\Large $\frac{A}{B}$} $\theta \, a \neq 0$, \qquad \quad  \ $v=0$,  \qquad \qquad \
$\nu \neq 0$, \qquad \qquad   $\dot{\mu} =0$ 
\\ \hline

C$_7$ & \phantom{\Large $\frac{A}{B}$} $\theta \, a \neq 0$, \qquad \quad \ $v=0$, \qquad \qquad \
$\nu = 0$, \qquad \qquad   $\omega \neq 0$ \qquad \quad \ \ \ 
  \\
\hline

 C$_8$ & \phantom{\Large $\frac{A}{B}$} $\theta \, a \neq 0$, \qquad \quad \  $v=0$,
\qquad \qquad \ $\nu = 0$, \qquad \qquad  $\omega = 0$ \phantom{hola}
\\
\hline \\[-5mm] \hline
\end{tabular}
}
%\end{ruledtabular}
\caption{The time-like unit vectors $u$ can be classified in eight classes C$_i$ ($i=1,...,8$) defined by relations imposed on the differential concomitants of $u$ given in (\ref{C-a}) (\ref{C-b}) (\ref{C-c}).}
\label{table-1}
\vspace{-5mm}
\end{table}

\begin{definition}
A time-like unit vector $u$ is said to be of class {\em C$_i$ ($i=1,...,8$)} if it satisfies the relations given in Table \ref{table-1}, where
\begin{eqnarray}
\theta = \nabla \cdot u \, , \quad  a \equiv i(u)\dif u \, , \quad \omega = *(u \wedge \dif u) \, ,  \label{C-a} \\
v = *(a \wedge \dif a)  \, , \quad  c = \frac{1}{a^2} i(a) \dif a  \, ,\label{C-b} \\
\mu  = \frac{(c,u)}{\theta} \, , \quad \nu =  2 + \frac{1}{\theta a^2} i(a) \dif \theta \, .  \label{C-c}
\end{eqnarray}
\end{definition}

Then, the result in the above section can then be summarized by the following two theorems:
\begin{theorem}
{\em ({\bf of characterization of CIG velocities})} A time-like unit vector $u$ of class {\em C$_i$ ($i=1,...,8$)}
is the velocity of a classical ideal gas if, and only if, it satisfies the differential system {\em S$_i$} given in Table \ref{table-2}, where
\begin{eqnarray}
f_1[u] \equiv -\theta \frac{(\omega, a)}{(v, u)} \, ;  \qquad \\  f_2[u] \equiv \frac{(x, v)}{v^2} \, , \quad  x = x[u] \equiv *(\dif \theta \wedge u \wedge a) \, ; \qquad  \\
f_{\pm}[u] \equiv \frac{1}{2 \dot{\mu}}[\theta \mu \nu - \dot{\nu} \pm \sqrt{\Delta}]  \, ,  \qquad \\[1mm]  \Delta = \Delta[u] \equiv  (\dot{\nu} - \theta \mu \nu)^2 - 4 \dot{\mu} \theta \nu (1- \nu)   \, ;  \qquad \\[1mm]
f_3[u] \equiv \frac{\theta \nu (1- \nu)}{\dot{\nu} - \theta \mu \nu} \, ;  \qquad \\
 f_4[u] \equiv - \frac{2 \theta^2}{\dot{\theta}} \, ;  \qquad  f_5[u] \equiv - \frac{\theta}{(c,u)}  \, .  \qquad
 \end{eqnarray}
\end{theorem}
%
%\newpage
%%
\begin{theorem}
The pairs $(\rho,p)$ of hydrodynamic quantities associated with a CIG velocity of class {\em C$_i$ ($i=1,...,8$)} are determined by relations {\em H$_i$} given in Table \ref{table-3}.
\end{theorem}
%\vspace{-6mm}
%
%
\begin{table}[h]
\vspace{-6mm}
%\begin{ruledtabular}
\normalsize{
%\vspace{1cm}
\noindent
\begin{tabular}{cl}
\hline \\[-5mm] \hline
 \phantom{\Large $ \quad \frac{A}{B}$} \qquad  \  & \hspace*{-4mm}  \ \ \ \  Necessary and sufficient conditions
   \\[1mm] \hline
S$_1$ & \phantom{\Large $\frac{A}{B}$}  \hspace*{-4mm}  $v = 0$, \qquad \ \
$i(u) \, \dif a =0 $
\\[1mm]
\hline

S$_2$  & \phantom{\Large $\frac{A}{B}$}  \hspace*{-4mm} $\dif u =0$, \qquad
$\dif \theta \wedge u =0 $
\\[1mm]
\hline

S$_3$  & \phantom{\Large $\frac{A}{B}$} \hspace*{-4mm} $\dif f \neq 0$,
\qquad (23),  \qquad  (24), \qquad with \ $f=f_1 [u] $
\\[1mm] \hline

S$_4$  & \phantom{\Large $\frac{A}{B}$} \hspace*{-4mm}  $\dif f \neq
0$, \qquad (23),  \qquad  (24), \qquad  with \ $f=f_2 [u] $ \\[1mm] \hline

S$_5$ & \phantom{\Large $\frac{A}{B}$}  \hspace*{-4mm}  $\Delta > 0$, \quad $\dif f \neq 0$,
\quad (23),  \quad  (24), \quad  with \ $f=f_{+} [u] \ \  $ or $ \ \ f = f_{-} [u] \  $ \quad
\\[1mm] \hline

 S$_6$ & \phantom{\Large $\frac{A}{B}$} \hspace*{-4mm}
 $\dot{\nu} - \theta \mu \nu \neq 0$, \quad $\dif f \neq 0$, \quad (23),  \quad  (24), \qquad
with \ $f=f_{3} [u]  \quad$
   \\[1mm]
\hline

S$_7$  &  \phantom{\Large $\frac{A}{B}$}  \hspace*{-4mm} $\dot{\theta}
\neq 0 $, \qquad \ \  (23),  \qquad  (24), \qquad  with  \  $f=f_{4} [u] $
\\[1mm] \hline

S$_8$  & \phantom{\Large $\frac{A}{B}$} \hspace*{-4mm} $(c,u) \neq 0 $,
 \quad (23),  \qquad  (24), \qquad with \ $f=f_{5} [u] $ \\[1mm] 
\hline \\[-5mm] \hline
\end{tabular}
}
%\end{ruledtabular}
\caption{The differential system S$_i$ gives the necessary and sufficient conditions for a unit vector of class C$_i$ to be the velocity of a classical ideal gas.}
\label{table-2}
\vspace{-9mm}
\end{table}
\begin{table}[h]
%\begin{ruledtabular}
\normalsize{
\begin{tabular}{cll}
\hline \\[-5mm] \hline
\\[-4mm]
&  \qquad  $p=p_0$, \qquad \quad $ \rho = \rho(\varphi_i) $;  \qquad \qquad \qquad
$ \dot{\varphi_i} =0 $ 
%\phantom{\large $\frac{A}{B}$} 
\qquad \qquad \quad    \
($a=0$) \quad
\\[-2.5mm]
%$\vdots$
\quad \quad  H$_1  $ \quad &   \\[-3mm]
&  \qquad    $p=p(\alpha)$, \qquad  $\rho = - p(\alpha)  - \beta\,  p^\prime (\alpha) $;   \quad \ \
$ \dif \alpha = \beta \, a $ \phantom{\Large $\frac{A}{B}$}   \qquad \quad
($a\not=0$) \ \ 
  \\[2mm] \hline
\quad \quad H$_2$ \quad & \qquad $p= C \, e^{\!  -\gamma \!  \int \!\theta(t) \dif t}$,  \quad
$\rho= \frac{p}{\gamma - 1}\! + \!B(\varphi_i) p^{\frac{1}{\gamma}}$;
\ \  \quad $\ u = - \dif t$, \ \quad  \quad $\dot{\varphi_i} =0$ \hspace*{-5mm} \phantom{\LARGE $\frac{A}{B}$} \\[2mm] \hline

\quad \quad H$_3$ \quad&  \phantom{\large $\frac{A}{B}$} \\[-1.5mm]
\quad \quad H$_4$  \quad 
%$\vdots$
&
%\phantom{\Large $\frac{A}{B}$}
 \\ [-2.5mm]
& \qquad   $p = C e^{\gamma \psi} $,  \qquad $\rho = p( \gamma f - 1)$; \qquad \qquad  $\dif
\psi = \theta u - f a $\\[-4.5mm]
\quad \quad H$_5$ \quad & \\[-1.mm]
\quad \quad H$_6$ \quad &   \\[0mm]  \hline \\[-4.5mm]
\quad \quad H$_7$ \quad &  \phantom{\normalsize $\frac{a}{b}$}  \hspace{7.7cm}  $\dif \psi = \theta u - f a, \, \, \beta a = \dif \alpha$
\\[-2.5mm]
%$\vdots$
&   \qquad   $p = C e^{\gamma \psi} $,  \qquad $\rho = p( \gamma f_{\varphi} - 1)$;\\[-2mm]
\quad \quad H$_8$  \quad &   \hspace{8.2cm} $f_{\varphi} = f + \beta \varphi(\alpha)$  \\[1mm]
\hline \\[-5mm] \hline
\end{tabular}
}
%\end{ruledtabular}
\caption{Row H$_i$ shows the hydrodynamic pairs $(\rho,p)$ associated with a CIG velocity of class C$_i$ by the inverse problem.}
%They can be obtained in terms of: (i) differential concomitants of $u$ like  $\theta$, $\gamma$ and $f$, (ii) functions like $alpha$, $\beta$, $t$, $\varphi$ and $\psi$, which   }
\label{table-3}
%\vspace{-1mm}
\end{table}
%
%
%\newpage
%

It is worth remarking that the results of theorems above offer an IDEAL characterization of the velocities of a CIG. This means that an algorithm can be built that allows us to distinguish every class C$_i$ and to test the labeling conditions S$_i$. We present this algorithm as a flow diagram (see below). The label in the top presents the seven concomitants of the unit vector $u$ that allow us to distinguish the different classes: $u$ itself, three first-order differential coefficients, $\theta$, $a$ and $\omega$, and three second-order differential concomitants, $v$, $\nu$ and $\mu$. Conditions in diamonds discriminate the different classes. If condition $n$ holds and the previous ones $n-1$ do not hold, then the velocity belongs to class C$_n$ and it must fulfill the necessary and sufficient conditions S$_n$ in order to be a CIG velocity.

\vspace{2.0mm}
\hspace*{-2cm}
 \setlength{\unitlength}{1.15cm} {\small 
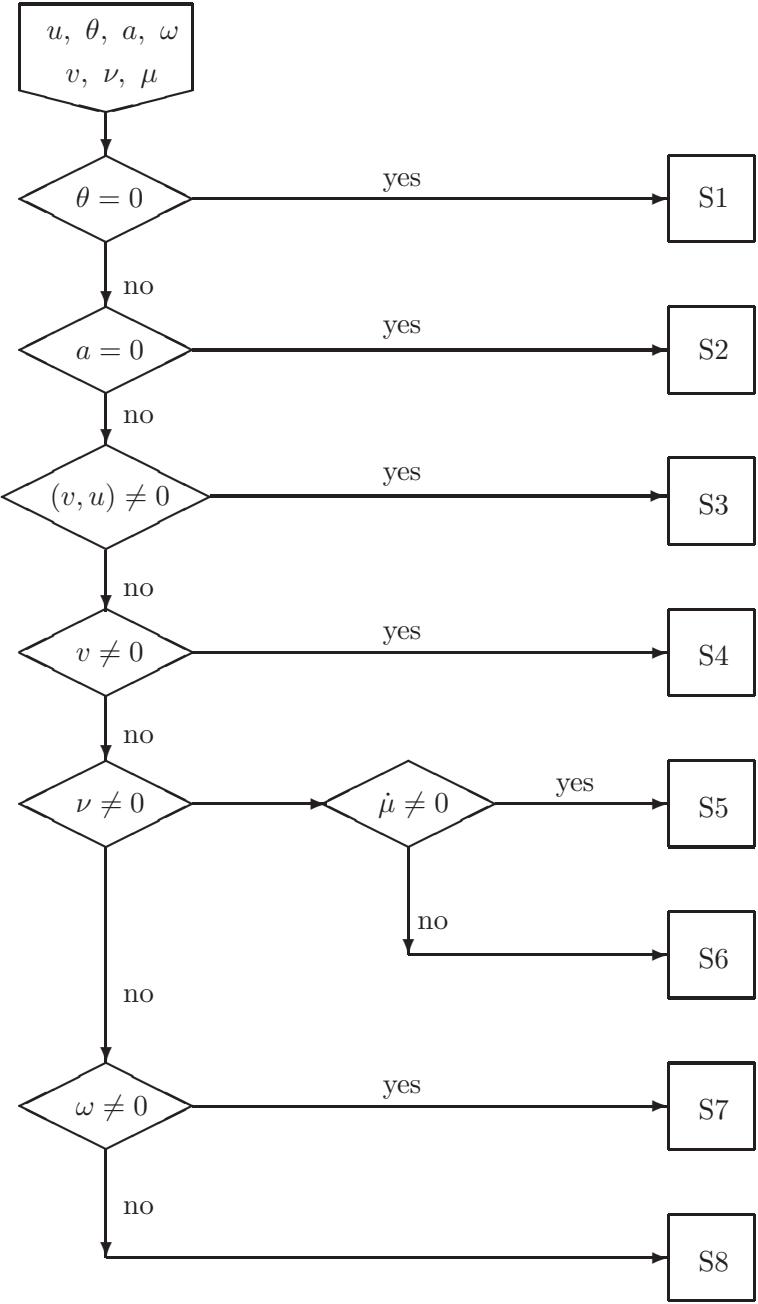
\noindent
\begin{picture}(0,18)
\thicklines

\put(4,17){\line(-4,-1){1}}
 \put(2,17){\line(4,-1){1}}
\put(2,17){\line(0,1){1}} \put(4,18){\line(-1,0){2}}
\put(4,18){\line(0,-1){1}} \put(2.2,17.6){$ \ u , \ \theta , \ a , \
\omega$}
 \put(2.2,17.1){$ \ \ \ v , \ \nu , \ \mu$}
%%%%%%%%%%%%%%%%
\put(3,16.75){\vector(0,-1){0.5}}

%%%%%%%%%%%%%%%%%%%%%%%
\put(3,16.25){\line(-2,-1){1}} \put(3,16.25){\line(2,-1){1}}
\put(3,15.25){\line(2,1){1}} \put(3,15.25){\line(-2,1){1}}
\put(2.65,15.65){$ \theta  = 0$}

%%%%%%%%%%%%%%%%%%%%%%%%%%%%%%%%%%%%%

\put(4,15.75){\vector(1,0){5.5}}

%%%%%%%%%%%%%%%%%%%%%%%

\put(9.5,15.25 ){\line(1,0){1}} \put(9.5,15.25){\line(0,1){1}}
\put(10.5,16.25){\line(-1,0){1}} \put(10.5,16.25){\line(0,-1){1}}

\put(9.85,15.65){S1}

%%%%%%%%%%%%%%%%%%%%%%%%%%
\put(3,15.25){\vector(0,-1){0.75}}
%%%%%%%%%%%%%%

\put(3,14.5){\line(-2,-1){1}} \put(3,14.5){\line(2,-1){1}}
\put(3,13.5){\line(2,1){1}} \put(3,13.5){\line(-2,1){1}}
\put(2.65,13.9){$a =0$}

%%%%%%%%%%%%%%%%%%%%%%%

\put(4,14){\vector(1,0){5.5}}

%%%%%%%%%%%%%%%%%%%%%%%
\put(9.5,13.5 ){\line(1,0){1}} \put(9.5,13.5){\line(0,1){1}}
\put(10.5,14.5){\line(-1,0){1}} \put(10.5,14.5){\line(0,-1){1}}

\put(9.85,13.9){S2}

%%%%%%%%%%%%%%%%%%%%%%%%%%
\put(3,12.9){\line(-2,-1){1.2}} \put(3,12.9){\line(2,-1){1.2}}
\put(3,11.7){\line(2,1){1.2}} \put(3,11.7){\line(-2,1){1.2}}

%\put(3,13){\line(-3,-1){2}} \put(3,13){\line(3,-1){2}}
%\put(3,11.65){\line(3,1){2}} \put(3,11.65){\line(-3,1){2}}
\put(2.35,12.2){$(v,u)  \neq 0$}

%%%%%%%%%%%%%5
\put(9.5,11.75 ){\line(1,0){1}} \put(9.5,11.75){\line(0,1){1}}
\put(10.5,12.75){\line(-1,0){1}} \put(10.5,12.75){\line(0,-1){1}}

\put(9.85,12.1){S3}

%%%%%%%%%%%%%%%%%%%%%%%%%%%%%%

\put(4.21,12.31){\vector(1,0){5.27}}

%%%%%%%%%%%%%%%%%%%%%%%%%%%%%%5

\put(3,13.5){\vector(0,-1){0.6}}

%%%%%%%%%%%%%%%%%%%%%%%%%

\put(9.5,10 ){\line(1,0){1}} \put(9.5,10){\line(0,1){1}}
\put(10.5,11){\line(-1,0){1}} \put(10.5,11){\line(0,-1){1}}

\put(9.85,10.35){S4}

%%%%%%%%%%%%%%%%%%%%%%%%%%%%%%

\put(3,11){\line(-2,-1){1}} \put(3,11){\line(2,-1){1}}
\put(3,10){\line(2,1){1}} \put(3,10){\line(-2,1){1}}
\put(2.65,10.4){$v \neq 0$}

%%%%%%%%%%%%%%%%%%%%%%
\put(3,11.68){\vector(0,-1){0.7}}

%%%%%%%%%%%%%%%%%%%%%%

\put(4,10.5){\vector(1,0){5.5}}

%%%%%%%%%%%%%%

\put(3,9.25){\line(-2,-1){1}} \put(3,9.25){\line(2,-1){1}}
\put(3,8.25){\line(2,1){1}} \put(3,8.25){\line(-2,1){1}}
\put(2.65,8.65){$\nu \neq 0$}

%%%%%%%%%%%%%%%%%%%%%%

\put(6.5,9.25){\line(-2,-1){1}} \put(6.5,9.25){\line(2,-1){1}}
\put(6.5,8.25){\line(2,1){1}} \put(6.5,8.25){\line(-2,1){1}}
\put(6.15,8.65){$\dot{\mu} \neq 0$}

%%%%%%%%%%%%%%%%%%%%%%

%%%%%%%%%%%%%%%%%%%%%%%%%

\put(9.5,8.25 ){\line(1,0){1}} \put(9.5,8.25){\line(0,1){1}}
\put(10.5,9.25){\line(-1,0){1}} \put(10.5,9.25){\line(0,-1){1}}

\put(9.85,8.6){S5}

%%%%%%%%%%%%%%%%%%%%%%%%%%%%%%

%%%%%%%%%%%%%%%%%%%%%%%%%

\put(9.5,6.5 ){\line(1,0){1}} \put(9.5,6.5){\line(0,1){1}}
\put(10.5,7.5){\line(-1,0){1}} \put(10.5,7.5){\line(0,-1){1}}

\put(9.85,6.85){S6}

%%%%%%%%%%%%%%%%%%%%%%%%%

\put(9.5,4.75 ){\line(1,0){1}} \put(9.5,4.75){\line(0,1){1}}
\put(10.5,5.75){\line(-1,0){1}} \put(10.5,5.75){\line(0,-1){1}}

\put(9.85,5.1){S7}

%%%%%%%%%%%%%%%%%%%%%%%%%%%%%%%%%%

\put(9.5,3 ){\line(1,0){1}} \put(9.5,3){\line(0,1){1}}
\put(10.5,4){\line(-1,0){1}} \put(10.5,4){\line(0,-1){1}}

\put(9.85,3.35){S8}

%%%%%%%%%%%%%%%%%%%%%%%%%%%%%%%%%%

%%%%%%%%%%%%%%%%%%%%%%

\put(3,5.75){\line(-2,-1){1}} \put(3,5.75){\line(2,-1){1}}
\put(3,4.75){\line(2,1){1}} \put(3,4.75){\line(-2,1){1}}
\put(2.65,5.15){$\omega \neq 0$}

%%%%%%%%%%%%%%%%%%%%%%

%%%%%%%%%%%%%%%%%%%%%%

\put(4,8.75){\vector(1,0){1.55}} \put(7.5,8.75){\vector(1,0){2}}

%%%%%%%%%%%%%%

\put(6.5,7){\vector(1,0){3}}

\put(6.5,8.25){\vector(0,-1){1.25}}

%%%%%%%%%%%%%%%%%%%%%%%%%%%%%%%%%%%%5
\put(4,5.25){\vector(1,0){5.5}}

%%%%%%%%%%%%%%%%%%%%%%%%%%%%%%%%%%%%5
\put(3,3.5){\vector(1,0){6.5}}

%%%%%%%%%%%%%%%%%%%%%%%%
\put(3,8.25){\vector(0,-1){2.5}}

\put(3,10){\vector(0,-1){0.75}}

\put(3,4.75){\vector(0,-1){1.25}}

%%%%%%%%%%%%%%%%%%%%%%%%%%%%%%%%%%%
%%%%%%%%%%%%%%%%%%%%%%%%%%%%%%%%%

 %%%%%%%%%%%%%%%%%%%%%%%%%%%%%%yes y no
 \put(6.2,10.65){yes}
  \put(6.2,12.5){yes}

\put(6.2,15.9){yes}
  \put(6.2,14.2){yes}

  \put(8.2,8.9){yes}
  \put(6.2,5.4){yes}

 %\put(3.2,16.35){no}

\put(3.2,14.65){no}

\put(3.2,13.15){no}

\put(3.2,11.15){no}

\put(3.2,9.45){no}

\put(3.2,6.45){no}

\put(3.2,4 ){no}

\put(6.6,7.3){no}

\end{picture} }
\vspace{-3.0cm}

\noindent
%{\small FLOW DIAGRAM. Algorithm that allows us to distinguish the eight classes of classical ideal gas velocities.}
%%%%%%%%%%%%%%%%%%%%%%

\section{Some classical ideal gas solutions}
\label{sec-CIG-examples}

In this section we apply our results to obtain test solutions to the hydrodynamic equation that model a classical ideal gas in local thermal equilibrium. We do not try to present an exhaustive analysis but just point out a method for the search of solutions in further work.

%%%%%%%%%%%%%%%%%%%%%%

\subsection{Classical ideal gas with a stationary flow}
\label{subsec-CIG-Killing}

Firstly we analyze a CIG with unit velocity $u$ such that $\xi = |\xi| u$ is a Killing vector. Then, if $\theta$ is the expansion, $\sigma$ the shear, and $a$ the acceleration of $u$, we have:
\be
\theta = 0 \, , \qquad \sigma =0 \, , \qquad  \dif a = 0 \, .
%  \, .
\ee
Moreover $a = \dif \alpha$, where $\alpha = \ln |\xi|$. Thus, the unit vector $u$ belongs to the class C$_1$ and it fulfills the necessary and sufficient conditions S$_1$. Consequently, if we take into account proposition \ref{propo-theta=0}, we obtain:
\begin{proposition} \label{propo-killing} 
Let $\xi$ be a time-like Killing vector, then $u = \xi/|\xi|$ is the unit velocity of any classical ideal gas.

When $a=0$, the pressure is an arbitrary constant, $p=p_0$, and the energy density is an arbitrary $u$-invariant function, $\rho = \rho(\varphi_i)$, $\dot{\varphi_i}=0$, $i=1,2,3$.

When $a \not=0$, the pressure is an arbitrary function of the norm of $\xi$, $p=p(\alpha)$, $\alpha = \ln |\xi|$, and the energy density is given by $\rho = \rho(\alpha) \equiv -p(\alpha) - p'(\alpha)$.
\end{proposition}

Note that we have, necessarily, a barotropic evolution, $d \rho \wedge d p=0$. Moreover, for any adiabatic index $\gamma$, the CIG thermodynamic scheme is given by the expressions delivered in lemma \ref{lemma-gic}.

The family of pairs $(\rho,p)$ that can be obtained in solving the inverse problem can be constrained by imposing some additional physical requirements. For example, we can consider a specific barotropic relation $p=p(\rho)$ derived from a particular evolution.

These results apply for static spherically symmetric space-times for both test fluids at rest in a given gravitational field and perfect fluid solutions of the Einstein equation. In \cite{CFS-CIG} we have analyzed the stellar structure equations for a self-gravitating classical ideal gas: (i) in thermal equilibrium with a non-vanishing conductivity, (ii) with an isothermal configuration and vanishing conductivity. These two physical situations have also been considered for a test classical ideal gas in the Schwarzschild space-time \cite{CFS-CIG}.

%%%%%%%%%%%%%%%%%%%%%%%

\subsection{Classical ideal gas with a conformally stationary flow}
\label{subsec-CIG-CK}

Let us consider now a CIG with unit velocity $u$ such that $\xi = |\xi| u$ is a conformal Killing vector. Then, we have:
\be \label{CK}
\dif \left(a - \frac13 \theta u\right) = 0 \, , \qquad \sigma =0   \, .
%  \, .
\ee
Moreover $a - \frac13 \theta u = \dif \alpha$, where $\alpha = \ln |\xi|$. Let us suppose that $\xi$ is not a Killing vector, that is, $\theta \not=0$.

When $a=0$, we have $\dif (\theta u) = 0$, and the unit vector $u$ belongs to class C$_2$ and it fulfills the necessary and sufficient conditions S$_2$. Consequently, if we take into account proposition \ref{propo-a=0}, we obtain:
\begin{proposition} \label{propo-CK-a=0}
Let $\xi$ be a time-like conformal Killing vector tangent to a geodesic congruence, then $u = \xi/|\xi|$ is the unit velocity of any classical ideal gas.

Moreover, a function $t$ exists such that relations {\em (\ref{a=0-t})} hold and, for any adiabatic index $\gamma$, the pressure and the energy density are given, respectively, by {\em (\ref{a=0-p})} and {\em (\ref{a=0-rho})}.
\end{proposition}

Otherwise, when $a \not=0$, (\ref{CK}) implies that pair $(u, f_0)$, with $f_0 = 3$, fulfills equations (\ref{fdot}) and (\ref{dfa}) for $\gamma = 4/3$. But it does not fulfill conditions in propositions \ref{propo-uf-gamma} or \ref{propo-uf} because $f$ is a constant. Consequently, a necessary condition for $u$ to be the velocity of a CIG is that the system  (\ref{fdot}, \ref{dfa}) admit another non-constant solution $f$. Then, proposition \ref{propo-dues-f} implies that $u$ necessarily belongs to classes C$_7$ or C$_8$, and from proposition \ref{propo-v=0-Y=0} we obtain:
\begin{proposition} \label{propo-CK-anot=0}
Let $\xi$ be a time-like conformal Killing vector which is not tangent to a geodesic congruence, then $u = \xi/|\xi|$ is the unit velocity of a classical ideal gas if, and only if, it fulfills $v=0$ and $\nu=0$, where $v$ and $\nu$ are given in {\em (\ref{C-b})} and {\em (\ref{C-c})} (it belongs to classes {\em C$_7$} and {\em C$_8$}). Moreover the adiabatic index is $\gamma=4/3$.

Then, two functions $\alpha, \beta$ exist such that $\beta a = \dif \alpha$ and the pressure and the energy density are given, respectively, by:
\be
p =  \frac{C}{|\xi|^4}\, , \qquad \rho = 3p [1+\frac43 \beta \varphi(\alpha)]  \, .
\ee
\end{proposition}

In the following subsections we apply propositions \ref{propo-CK-a=0} and \ref{propo-CK-anot=0} to obtain CIG test solutions: (i) at rest with respect the cosmological observer in an arbitrary Friedmann-Lema\^itre-Robertson-Walker (FLRW) universe, and (ii) with a radial conformally stationary flow in Minkowski space-time.

%%%%%%%%%%%%%%%%%%%%%%%

\subsection{Comoving classical ideal gas in FLRW universes}
\label{subsec-CIG-FLRW}

The FLRW universes are perfect
fluid solutions of Einstein equations with line element:
\be \label{metric-FLRW} \dif s^2 = - \dif t^2 + \frac{R^2(t)}{\left[1+ \frac14
\varepsilon r^2\right]^2}(\dif r^2 + r^2 \dif \Omega^2)  \,   ,
\ee
with $\varepsilon = 0, 1,\! -1$. The cosmological observer $u= - \dif t$ defines a geodesic conformally stationary flow and, consequently, proposition  \ref{propo-CK-a=0} applies. Then, taking into account that $\theta = 3 \dot{R}/R$, and the expressions (\ref{a=0-p}) and (\ref{a=0-rho}) for the pressure and energy density and the  expressions in lemma \ref{lemma-gic} for the thermodynamic scheme, we obtain:
\begin{proposition} \label{propo-FLRW}
In any FLRW universe a test solution of the fundamental system of the CIG hydrodynamics exists which is comoving with the cosmological observer.

Moreover, any adiabatic index $\gamma$ is possible, and the energy density, the pressure, the matter density and the temperature depend on the expansion factor $R(t)$ as:
\begin{eqnarray}
\rho(R) = N(x^i) \left(\frac{R_0}{R}\right)^3 + \frac{p_0}{\gamma-1}
\left(\frac{R_0}{R}\right)^{3\gamma}   , \qquad  \label{FLRW-ideal} \\
 \label{FLRW-ideal-p} p(R) = p_0 \left(\frac{R_0}{R}\right)^{3
\gamma} , \quad n(R) = N(x^i) \left(\frac{R_0}{R}\right)^{3} , \quad  \\
\label{FLRW-ideal-T} \Theta(R) = \frac{p_0}{k N(x^i)}
\left(\frac{R_0}{R}\right)^{3 (\gamma - 1)}  ,  \qquad \qquad \quad
\end{eqnarray}
where $N(x^i)$ is an arbitrary function of the spatial coordinates $r, \vartheta, \varphi$.
\end{proposition}
Note that we have homogeneous pressure and inhomogeneous energy density, matter density and temperature. We can consider CIG homogeneous models by taking $N(x^i) = n_0$.

Of course, the models in proposition above are test solutions in any FLRW universe. Then, a question naturally arises: can these CIG solutions be the source of the Friedmann model? The answer is affirmative. Indeed, we can pose the generalized Friedmann equation:
\begin{equation} \label{Friedmann-eq}
\frac{3 \dot{R}^2}{R^2}  +  \frac{3 \varepsilon}{R^2} \equiv \rho(R) \, ,
\end{equation}
for the function $\rho(R)$ given in (\ref{FLRW-ideal}), with $N(x^i) = n_0$. Then, the corresponding FLRW models have a pressure that takes the expression (\ref{FLRW-ideal-p}) and, as it has been pointed out in a previous paper \cite{CFS-CIG}, they can be interpreted as a self-gravitating GIG in isentropic evolution and with matter density and temperature given in (\ref{FLRW-ideal-p}) and (\ref{FLRW-ideal-T}).

%%%%%%%%%%%%%%%%%%%%%%%

\subsection{Classical ideal gas with a radial conformally stationary flow in Minkowski space-time}
\label{subsec-CIG-CK-Minko}

The conformally stationary motions in a flat or a conformally flat space-time have been widely analyzed in the literature (see, for example  \cite{AliciaJA} \cite{AliciaJAb}). In the space-time regions where we have a  time-like conformal Killing vector $\xi$ the results in subsection \ref{subsec-CIG-CK} apply.

When the flow is geodesic, $a =0$, then $u = \xi/|\xi|$ is the unit velocity of any CIG as a consequence of proposition \ref{propo-CK-a=0}. Moreover, it can easily be proved \cite{AliciaJA} that it corresponds to a Milne's observer. It is known \cite{rindler} that in coordinates adapted to the Milne's observer, $u = - \dif \tau$, the Minkowski metric can be written as a FLRW metric with cosmological time $\tau$, curvature $k=-1$ and expansion factor $R(\tau)=\tau$. Then, the CIG associated with $u$ as a consequence of proposition \ref{propo-CK-a=0} are precisely those presented in proposition \ref{propo-FLRW} for the particular case of the Milne universe. Moreover, we recover the known \cite{rindler} expression of the cosmological Milne time $\tau$ in terms of the spherical inertial coordinates $(t,r,\vartheta, \varphi)$:
\be
\tau = \frac{t}{\sqrt{t^2-r^2}} \, , \qquad  t>r \, .
\ee

When $a\not=0$, if $u$ is the velocity of a CIG, then it belongs to classes C$_7$ or C$_8$ as a consequence of  proposition \ref{propo-CK-anot=0}. On the other hand, from the Ricci identities for $u$ and conditions (\ref{CK}) one has $\dif \theta \wedge u =0$ and we obtain the value $\nu = 2 \not=0$ for the scalar $\nu$ given in (\ref{C-c}). Consequently, neither class C$_7$ nor class C$_8$ are possible.

We can summarize the results in this subsection as follows.
\begin{proposition} \label{propo-CK-Minko}
In Minkowski space-time the only expanding radial conformally stationary congruences which define the flow of a classical ideal gas are the Milne's ones.
\end{proposition}
%%

%%%%%%%%%%%%%%%%%%%%%%%

\subsection{Classical ideal gas in geodesic radial motion in Minkowski space-time}
\label{subsec-CIG-Minko-a=0}

In the Minkowski space-time a radial time-like unit covector $u$ takes the following expression in inertial spherical coordinates $(t,r, \vartheta, \varphi)$:
\be
u = - \cosh \phi \, \dif t + \sinh \phi \, \dif r \, , \quad \phi = \phi(t,r, \vartheta, \varphi) \, .
\ee
When $u$ is geodesic, $a=0$, then it is the velocity of a CIG if, and only if, $\dif u= 0$ and $\dif \theta \wedge u = 0$ as a consequence of proposition \ref{propo-a=0}. Then, the hyperbolic angle $\phi$ does not depend on the angular coordinates: $\phi = \phi(t,r)$. Moreover, the acceleration $a$ and the expansion $\theta$ take, respectively, the expressions:
\begin{eqnarray}
a= (\cosh \phi  \, \phi_{ t}\! + \! \sinh \phi \, \phi_r) (\! - \! \sinh \phi \, \dif t \!+ \! \cosh \phi \, \dif r)  ,  \qquad \\[1mm]
\theta =  \sinh \phi  \, \phi_{ t} + \cosh \phi \, \phi_r +\frac{2}{r} \sinh \phi  .  \qquad
\end{eqnarray}
By using these expressions we can impose $a=0$ and $\dif \theta \wedge u =0$ and we arrive to a differential system whose only solution is $\tanh \phi = r/(t-t_0)$, which corresponds to a Milne observer. Alternatively, a straightforward calculation shows that $\dot{\theta} + \theta/3 = 0$, and then the Ricci identities lead to a vanishing shear, $\sigma =0$. Consequently, we have a radial conformally stationary flow and proposition \ref{propo-CK-Minko} applies. Thus, we arrive to the following.
\begin{proposition} \label{propo-a=0-Minko}
In Minkowski space-time the only expanding radial geodesic congruences which define the flow of a classical ideal gas are the Milne's ones.
\end{proposition}
%%
%

%%%%%%%%%%%%%%%%%%%%%%%

\ack 
We are grateful to B. Coll for having instilled into us the conceptions that motivate this work and other related research. We would like to thank J. A. Morales-Lladosa for his suggestions and comments. This work has been supported by the Spanish ``Ministerio de Econom\'{\i}a y Competitividad", MICINN-FEDER project FIS2015-64552-P.

%%%%%%%%%%%%%%%%%%%%%%%

\end{document}